\documentclass[number, preprint, sort&compress, 1p]{elsarticle}
\pdfoutput=1
\usepackage{graphicx}
\usepackage[english, american, ngerman]{babel}
\usepackage{lmodern}

\usepackage[utf8]{inputenc}
\usepackage[T1]{fontenc}
\usepackage{amsmath,amssymb}   
\usepackage{booktabs}
\usepackage{import}
\usepackage{units}
\usepackage{tabularx, tabulary}
\usepackage{ragged2e}
\usepackage[colorlinks=true, allcolors=black, bookmarksnumbered=true]{hyperref}
\usepackage{bookmark}
\usepackage{xcolor}
\usepackage{colortbl}
\usepackage{subcaption}
\usepackage{longtable}
\usepackage{relsize}
\usepackage{pgfplots}
\usepackage{dsfont}

\pgfplotsset{compat=newest}
\pgfplotsset{plot coordinates/math parser=false}
\newlength\figureheight
\newlength\figurewidth 

\let\vec\mathbf

\newcommand{\R}{\mathbb{R}} 																			
\usepackage{hyperref}
\renewcommand{\vec}[1]{\boldsymbol{#1}}
\DeclareMathOperator{\diag}{diag}

\usepackage{amsfonts}
\newcommand{\overbar}[1]{\mkern 1.5mu\overline{\mkern-1.5mu#1\mkern-1.5mu}\mkern 1.5mu}

\begin{document}
\selectlanguage{american}

\begin{frontmatter}

\title{Self-Tuning State Estimation for Adaptive Truss Structures Using Strain Gauges and Camera-Based Position Measurements\tnoteref{copyright}\tnoteref{doi}}
\tnotetext[copyright]{\copyright~2020. This manuscript version is made available under the CC-BY-NC-ND 4.0 license \url{http://creativecommons.org/licenses/by-nc-nd/4.0/}}   
\tnotetext[doi]{Published version: \url{https://doi.org/10.1016/j.ymssp.2020.106822}}
\author[isys]{Alexander Warsewa\corref{cor}}
\ead{warsewa@isys.uni-stuttgart.de}
\author[isys]{Michael Böhm}
\author[ito]{Flavio Guerra}
\author[isys]{Julia Wagner}
\author[ito]{Tobias Haist}
\author[isys]{Cristina Tarín}
\author[isys]{Oliver Sawodny}
\cortext[cor]{Corresponding author.}
\address[isys]{Institute for System Dynamics, University of Stuttgart, Waldburgstraße 17/19, 70563 Stuttgart, Germany}
\address[ito]{Institute of Applied Optics, University of Stuttgart, Pfaffenwaldring 9, 70569 Stuttgart, Germany}

\begin{abstract}
In the context of control of smart structures, we present an approach for state estimation of adaptive buildings with active load-bearing elements. For obtaining information on structural deformation, a system composed of a digital camera and optical emitters affixed to selected nodal points is introduced as a complement to conventional strain gauge sensors. Sensor fusion for this novel combination of sensors is carried out using a Kalman filter that operates on a reduced-order structure model obtained by modal analysis. Signal delay caused by image processing is compensated for by an out-of-sequence measurement update which provides for a flexible and modular estimation algorithm. Since the camera system is very precise, a self-tuning algorithm that adjusts model along with observer parameters is introduced to reduce discrepancy between system dynamic model and actual structural behavior. We further employ optimal sensor placement to limit the number of sensors to be placed on a given structure and examine the impact on estimation accuracy. A laboratory scale model of an adaptive high-rise with actuated columns and diagonal bracings is used for experimental demonstration of the proposed estimation scheme.
\end{abstract}


\end{frontmatter}

\section{Introduction}
Equipped with active load-bearing elements, adaptive structures can react to different load cases and disturbances. This enables the construction of lightweight structures resulting in a significant decrease of embodied energy. Senatore et al.\,\cite{senatore2017shape} recently presented a prototype of an adaptive bridge. Control of a smart shell structure with actuated supports is demonstrated in\,\cite{heidingsfeld2015actuator}. Adaptivity allows those structures to be less massive and stiff than otherwise admissible. Here, we consider the state estimation problem for tall and lightweight adaptive buildings. Experimental validation of the proposed sensor systems and methods is conducted on a laboratory scale prototype. 

The control of adaptive structures requires precise knowledge of the state in real-time. Since it is not feasible to measure the entire state of a complex structure, it needs to be reconstructed from measurements. These are usually obtained from sensors spanning multiple domains and sampled at different rates. In addition to conventional strain gauge sensors, we employ a system comprised of a digital camera and emitters attached to the structures nodal points for displacement monitoring. Optical sensors such as laser Doppler vibrometers (LDVs) or fiber-Bragg sensors are often used in structural health monitoring; see \,\cite{aoyama2018vibration, kang2007estimation, kim2017dynamic, park20153d} for a range of examples. Our approach is similar to a motion capture system but uses potentially fewer cameras, can be realized with cost-effective hardware and is easily expandable and reusable. Due to the high precision of the optical measurements, the camera system is also suitable for structural parameter estimation.

Numerous methods for state and parameter estimation of mechanical structures have been proposed. An early review on state estimation in structural dynamics was presented by Waller and Schmidt in\,\cite{waller1990application}. They introduced the concept of modal observers which operate on systems of smaller order obtained by modal analysis of finite element (FE) models. Recently, Bayesian approaches to handle model uncertainties such as the Kalman and particle filter have become widely-used\,\cite{ching2006bayesian}. Smyth and Wu\,\cite{smyth2007multi} presented a modification of the Kalman filter to account for data sampled at different rates. The time update step is performed with the fast accelerometer measurements acting as system input, whereas the measurement update step is only executed when displacement data becomes available. Lourens et al.\,\cite{lourens2012joint} addressed the problem of joint input and state estimation for mechanical structures when the input forces are unknown. Their algorithm is also based on the Kalman filter, except that the input forces are replaced by optimal estimates. Azam et al.\,\cite{azam2015dual} introduced a dual implementation of the Kalman filter where the input forces are estimated in a first stage and subsequently used in the second stage for state estimation. Hernandez and Bernal\,\cite{hernandez2008state} derived a model-based observer for structural dynamics assuming that the primary source of error is a mismatch of the real system and the state-space matrices used to represent it. While considerably simpler to implement than Kalman-based observers, it is only applicable when displacements or their temporal derivatives are measured. 

For the sensor fusion of camera and strain gauge data, we focus on practical aspects. Large spatial distance between individual sensors, different sampling speeds or transmission delays often give rise to concern. We use Kalman filters for sensor fusion augmented by an out-of-sequence measurement (OOSM) update step as proposed by Bar-Shalom\,\cite{bar2002update}. This enables the processing of sensor data from the strain gauges at a higher sampling rate than the optical measurements and the handling of delays caused by image processing in the camera system.

The remainder of this contribution is divided into seven sections. First, the adaptive structures test bench developed in our laboratory is introduced in detail with a focus on the sensors used for observer design. In Sec.\,\ref{sec:modeling}, the dynamic model of the structure is derived. Modal analysis is employed to obtain a state space representation of our system which is suitable for real-time estimation and control. Based on previous contributions, Sec.\,\ref{sec:sensor_placement} briefly covers the problem of sensor placement for the test bench. Subsequently, the sensor fusion approach composed of a Kalman filter with OOSM update is presented. Sec.\,\ref{sec:parameter_tuning} introduces a simple optimization-based method for self-tuning of both filter and model parameters. Results are discussed in Sec.\,\ref{sec:results} followed by concluding remarks in Sec.\,\ref{sec:conclusion}.

\section{Adaptive Structures Test Bench}\label{sec:materials_and_methods}

\begin{figure}
 \centering
 \def\svgwidth{\textwidth}
 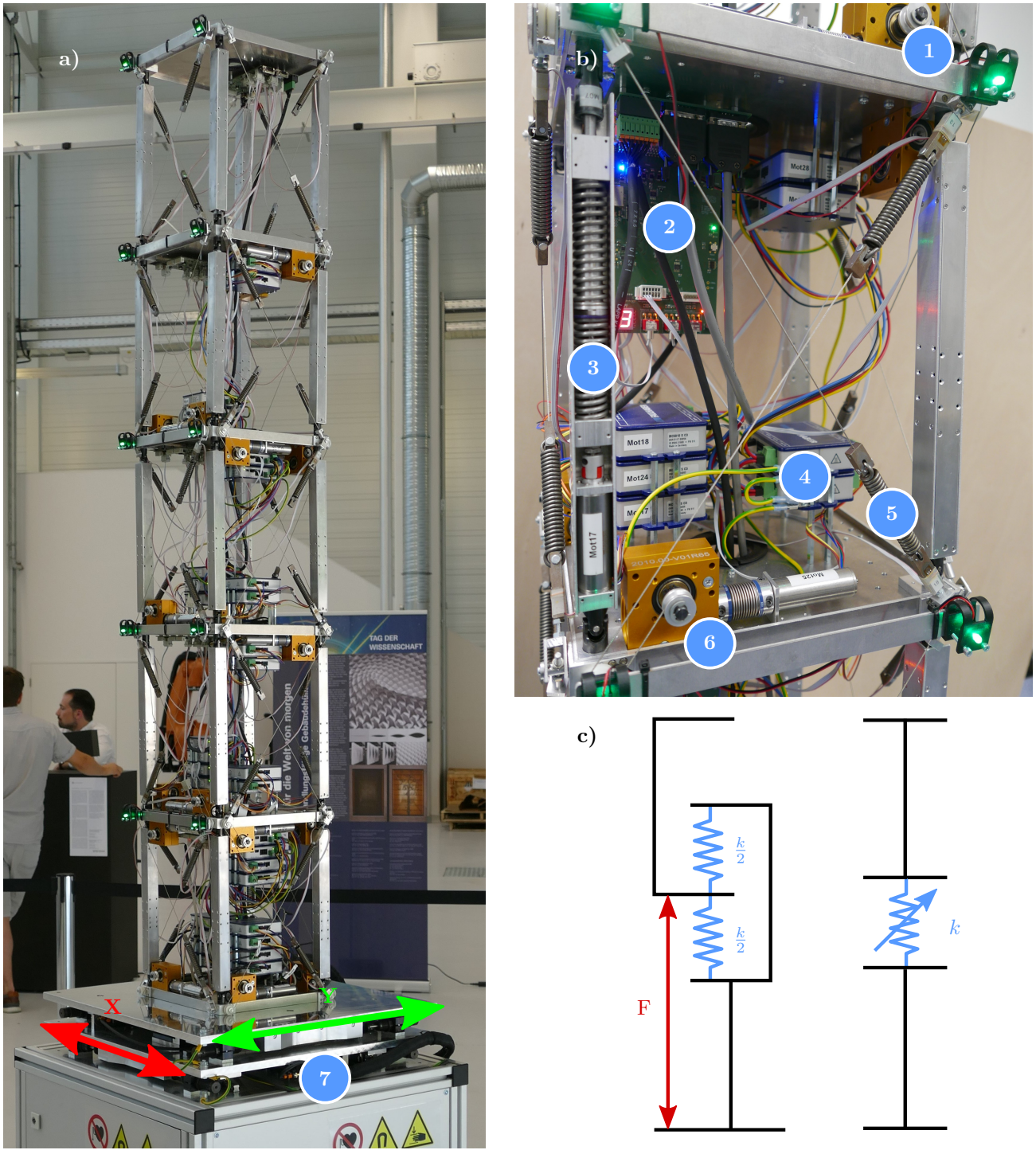 
 \caption[Adaptive Structures Test Bench]{Adaptive structures test bench. a)~View of all 5 modules including the controllable table (7) used for exciting the structure. b)~Closer view of the 3rd module showing LED emitters (1), custom signal processing hardware (2), an actuated column with dismantled cover (3), motor controllers (4) for active elements, passive diagonal bracings (5) and actuated bracings (6). c)~Parallel force actuation of columns (left) and change of pretension/elongation for bracings (right). }
 \label{fig:msm_real}
\end{figure}
An adaptive truss structure high-rise is being constructed on the campus of the University of Stuttgart in the scope of the collaborative research center CRC1244\,\cite{weidner2018implementation}. It is intended for demonstrating the concept of actuated truss structures to a larger audience. The test bench, which is introduced in the following, represents a scale 1:18 version of this demonstrator building. As the name implies, it is used to validate control engineering methods before they are applied to the actual high-rise demonstrator. The scaled building is designed to have eigenfrequencies similar to the ones expected for the demonstrator. Thus, only the dimensions are adopted to scale, while material stiffness and truss diameters are not. 

Fig.\,\ref{fig:msm_real}\,a) shows a full view of the scale model. Composed of 5 modules of identical size with a footprint of $26 \times 26\,\text{cm}$ and a height of $40\,\text{cm}$ each, it extends to a total height of $2\,\text{m}$ which corresponds to $36\,\text{m}$ in reality. Each module represents two stories of the demonstrator building. Intermediate floor plates were not installed to make room for the control hardware. A module comprises 4 vertical columns, 8 diagonal bracings and a ceiling plate. Columns are connected to plates via universal joints with two rotational DOFs about the local $x$- and $y$-axes. Rotation about the local $z$-axis is blocked. Bracings are attached to the structure via a rotational joint with one DOF at each end. The lowermost module is mounted to a controllable table (7) which can move in a plane parallel to the ground. Using the table, excitation of the structure as well as the simulation of earthquake primary waves is possible. Selected columns and bracings of the scale model are actuated, where the number of active elements differs in each module. The top module is entirely passive. 
A closeup view of the 3rd module in Fig.\,\ref{fig:msm_real}\,b) reveals how columns and bracings are actuated. When one of the panels of a column is removed (3), two identical springs are visible inside, which are connected in parallel. For actuated columns, the connecting point of the two springs can be moved by a motor via a threaded bar. This generates a force, as illustrated by the schematic drawing on the left of Fig.\,\ref{fig:msm_real}\,c). The stiffness of a diagonal bracing is determined by a spring which, in case of a passive bracing (5), is tautened between corner nodes using a steel cable and a screw system. One cable end of an active bracing (6) is attached to a guide roller which can be rotated using a DC-motor in combination with a worm gear. Accordingly, the spring of an active bracing can be directly extended or relaxed. In order to minimize slackening, all diagonal bracing springs are pretensioned. Each motor is driven by a motor controller (4) which in turn is controlled by custom signal processing hardware (2). Details on the test bench's computational hardware, sensors, actuators and data processing are given in the following.

\subsection{Signal Processing and Control of Actuated Components}

\begin{figure}
 \centering
 \def\svgwidth{\textwidth}
 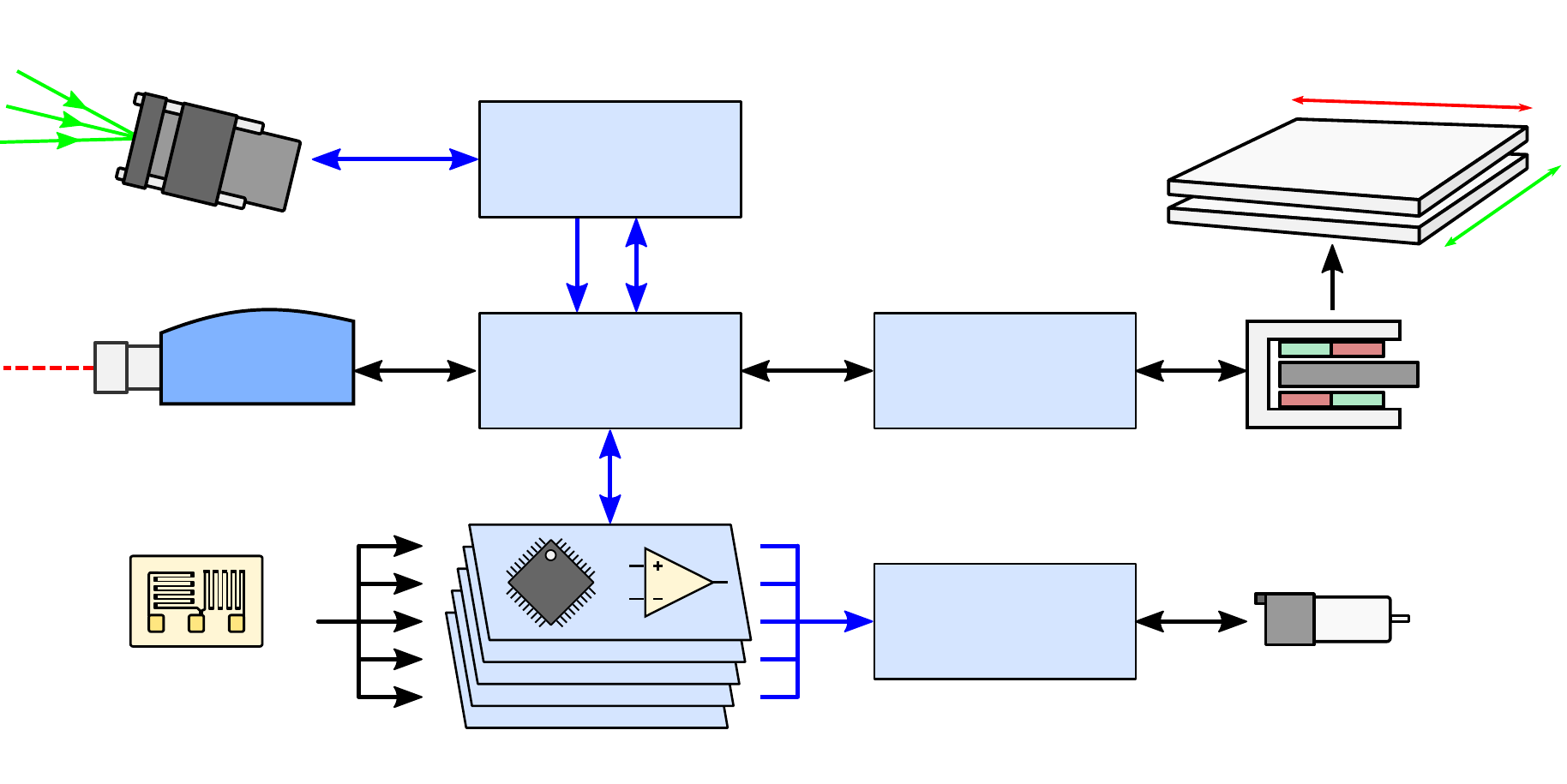 
 \caption{Interconnection of sensors, processing units and actuators for the adaptive structures test bench. Sensor signals from camera, vibrometer and strain gauges are processed on the host PC, a rapid prototyping device (MicroLabBox) and custom PCBs respectively. The MicroLabBox acts as CPU and controls table motion via intermediary servo controllers. The DC-motors of active truss elements are accessed by the signal processing boards story-wise.}
 \label{fig:hardware_overview}
\end{figure}

Fig.\,\ref{fig:hardware_overview} depicts the interconnection of sensors, actuators and computational hardware. A rapid prototyping device (dSpace MicroLabBox DS1202) acts as central processing unit. It is interfaced with the host PC via Ethernet and programmed using a Matlab/Simulink real-time interface in combination with dSpace ControlDesk. 

For measuring forces, each column and bracing is equipped with two half bridge strain gauges (Vishay EA-06-062TZ-350) mounted on opposing sides and connected to form a full bridge circuit. The sensors are attached to steel components with high flexural rigidity such that forces due to bending can be neglected. Strain gauge signals are amplified and processed on custom-built printed circuit boards (PCBs) that communicate with the MicroLabBox via CAN bus. Each story is equipped with one PCB stack that is entrusted with the task of reading out local sensors and controlling the actuated elements in the respective module. The microcontrollers on the PCBs can also be used for decentralized preprocessing and control. Motor controllers (Faulhaber MC5010) are accessed via local CAN buses and brushless DC-motors (Faulhaber 2264W024BP4 3692) are used for active truss elements. The $x$/$y$-motion of the table is controlled by servo amplifiers (Metronix ARS 2105) connected to analog I/O-ports of the MicroLabBox. Iron-free linear motors (Tecnotion UX 9N) in combination with optical encoders (Heidenhain LC485) facilitate the table motion. 

A digital camera (Ximea MC023MG-SY) attached to a tripod is located at a distance of approximately $2\,\text{m}$ from the test bench. It tracks the position of a total of 10 light emitting diodes (LEDs), positioned, as shown in Fig.\,\ref{fig:msm_real}\,a) and Fig.\,\ref{fig:msm_real}\,b) (1), in line with the nodal points of one face of the structure. Standard green LEDs with a diameter of $5\,\text{mm}$ are used. They are clipped to aluminum slats parallel to the plate edges with custom made 3D printed plastic brackets. The bracket angle can be adjusted such that the emitters point towards the camera. An optical bandwidth filter (ThorLabs FL05532-10) is mounted in front of the camera. Given the exposure time is sufficiently short, it passes only the green LED light. Image data is read in via USB and processed on the host PC. More details on how the emitter positions are determined are provided in the following section. 

Besides the strain gauges, three LDVs are connected to the MicroLabBox and provide accurate displacement information in one axis. They are used as a reference to validate the estimation results in Sec.\,\ref{sec:results}.

\subsection{Nodal Position Tracking}\label{sec:tracking}
Light from the LEDs attached to the scale model is registered by charge-coupled devices (CCDs) in the camera and needs to be mapped to nodal positions from pixel coordinates. Precise locations are usually obtained by high integration times which, however, result in low sampling frequencies. Mounting a filter that matches with the emitter wavelength in front of the camera lens significantly increases the signal-to-noise ratio (SNR) and allows for shorter sampling periods. 

To reconstruct the position of a each tracked point, a commonly used technique is to calculate the center of gravity (COG) of the corresponding gray values of the image spot. But this widespread technique has physical limitations which are, amongst other things, lens distortion, discretization errors and electron as well as photon noise. While the latter limitations cannot be easily overcome, lens distortion is compensated for with the help of a simple model. The algorithms used are part of the Open Source Computer Vision Library (openCV)\,\cite{opencv_library}. Correction of lens distortion is performed as follows
\[
\begin{bmatrix}
v' \\
w'
\end{bmatrix}
=
\begin{bmatrix}
v \left( 1 + k_1 r^2 + k_2 r^4 + k_3 r^6 \right) + 2 p_1 v w + p_2 \left( r^2 + 2 v^2 \right) \\
w \left( 1 + k_1 r^2 + k_2 r^4 + k_3 r^6 \right) + 2 p_2 v w + p_1 \left( r^2 + 2 w^2 \right)
\end{bmatrix},
\]
where $(v, w)$ are the image coordinates and $r^2 = v^2 + w^2$. The image point coordinates are corrected with radial distortion parameters $k_n$ and their tangential counterparts $p_n$ using a polynomial fit. In order to obtain an accurate mapping between object space and image space, the calibration process must involve an object of known form and dimensions. Printing a check pattern of known square size on a flat surface, the object-positions $(X, Y, Z)$ of the edges of the squares can be mapped to their image positions. This image sensor coordinate transformation from 3D-coordinates is done with two transformation matrices
\[
\begin{bmatrix}
v \\
w \\
1
\end{bmatrix}
=
\underbrace{
\begin{bmatrix}
f_x & 0 & c_x \\
0 & f_x & c_y \\
0 & 0 & 1
\end{bmatrix}
}_\text{intrinsic}
\underbrace{
\begin{bmatrix}
r_{11} & r_{12} & r_{13} & t_1 \\
r_{21} & r_{22} & r_{23} & t_2 \\
r_{31} & r_{32} & r_{33} & t_3 \\
\end{bmatrix}
}_\text{extrinsic}
\begin{bmatrix}
X \\
Y \\
Z \\
1
\end{bmatrix}.
\]
The left matrix describes intrinsically how much the image scales (focal lengths $f_x$ and $f_y$) and translates ($c_x$ and $c_y$). Extrinsic parameters, namely the camera orientation and its rotation with respect to the object, are entries of the right matrix. Obtained image positions of the emitters are preprocessed on the host PC, as shown in Fig.\,\ref{fig:hardware_overview}. A simple model comprising the expected number of points as well as their approximate location is used to test for measurement errors and to ensure that the information is always sent to the MicroLabBox in a predetermined order.

In the following sections, the model used for state estimation and the corresponding sensor fusion algorithm for the integration of camera system measurements and strain gauge data are presented.

\section{Modeling and Model Order Reduction}\label{sec:modeling}
\begin{figure}
	\centering
	\begin{minipage}{0.5\textwidth}
		\centering
		\includegraphics{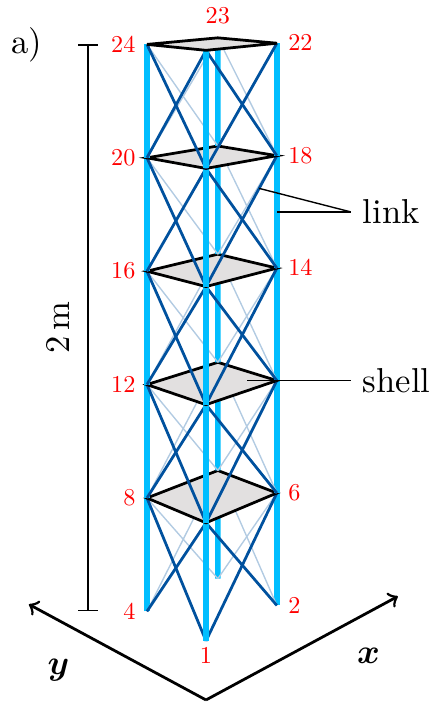}
	\end{minipage}
	\hfill
	\begin{minipage}{0.45\textwidth}
		\includegraphics{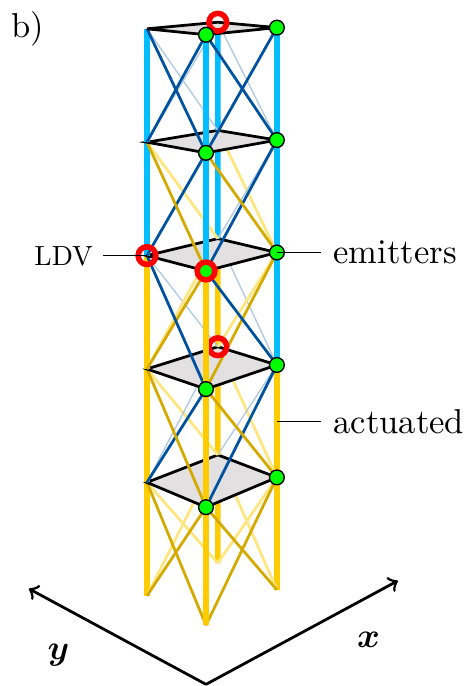}
	\end{minipage}
	\caption[Adaptive Truss Structure Scale Model]{Adaptive truss structure scale model. a) Columns and diagonal bracings are modeled in ANSYS as link elements, module floor and ceiling plates as shell elements. Node numbers are shown in red. b) Actuated elements (yellow), nodes tracked by the camera (green) and nodes measured by LDVs (red).}%
	\label{fig:msm_model}%
\end{figure}
For the application of engineering methods, a model with finite number of degrees of freedom (DOFs) is required. Computational load and accordingly hardware cost for model-based estimation and control scales with the model's complexity. The scale model dynamics can be described by the following linear second-order differential equation
\begin{align}
\begin{split}
	&\vec M \ddot{\vec q}(t) + \vec D \dot{\vec q}(t) + \vec K \vec q(t) = \vec F_a \vec u(t) + \vec E_\text{d} \vec z(t) , \quad t>0\\
	&\vec q(0) = \vec q_0, \quad \dot{\vec q}(0) = \vec q_1,
	\label{eq:system_dynamics}
\end{split}
\end{align}
where \(\vec q \in \R^n\) are the nodal DOFs, \(\vec M \in \R^{n \times n}\) is the mass matrix, \(\vec D \in \R^{n \times n}\) the damping matrix and \(\vec K \in \R^{n \times n}\) the stiffness matrix. \(\vec F_a \in \R^{n \times m_u}\) represents the input matrix for forces $\vec u(t)$ exerted by the actuators, where $m_u$ is the number of actuators. The matrix $\vec E_\text{d} \in \R^{n \times m_d}$ maps deadweight and disturbances $\vec z(t) \in \mathbb R^{m_d}$, such as wind- and payloads, to the model's DOFs. Accurate determination of the matrix \(\vec D\) is demanding, due to the complex interplay of different damping effects. A common approximation is the assumption of Rayleigh damping with 
\begin{align}
	\vec D = \alpha_0 \vec M + \alpha_1 \vec K.
\end{align}
Coefficients \(\alpha_0\) and \(\alpha_1\) can be determined through an experiment or by parameter identification as in\,\cite{ling2004element}. Rayleigh damping is not physically motivated, but it has several convenient properties. For instance, it ensures that the mode shapes remain the same as for the undamped system. This is desirable when performing a modal analysis of \,\eqref{eq:system_dynamics} to obtain the eigenmodes of the system. 

Mass and stiffness matrix of the structure are calculated using a FE model with element types as indicated in Fig.\,\ref{fig:msm_model}\,a). Columns and diagonal bracings are modeled as rods, module floors and ceilings as shells. Translational DOFs for nodes one to four are eliminated. Different stiffness values are assumed for active and passive columns due to constructional differences. In Fig.\,\ref{fig:msm_model}\,b) the 31 actuated elements are highlighted in yellow. For the scale model considered in this contribution, the number of DOFs in\,\eqref{eq:system_dynamics} amounts to $n = 120$, which can be prohibitively high for control applications. We employ a model order reduction based on modal analysis, where higher order eigenmodes are truncated\,\cite{gawronski2004advanced}. This is a widespread technique, as a few low-frequency modes often describe the system's behavior sufficiently precise. A modal truncation of the system is even more reasonable when considering the limited bandwidth of actuators and sensors. Only modes within the respective bandwidths can be properly controlled or observed. 

The generalized eigenvalue problem of the undamped system is formulated by plugging in the harmonic ansatz \(\vec q(t) = \boldsymbol{\varphi}_i e^{j\omega_i t}\) with
\begin{align}
(\vec K - \omega_i^2 \vec M)\boldsymbol \varphi_i = 0, \quad i = 1, 2, \dots n.
\end{align}
The eigenfrequencies are denoted by \(\omega_i\) and the eigenvectors by \(\boldsymbol \varphi_i\). Since the eigenvectors are not unique, we follow the common approach of choosing the ones that satisfy \(\boldsymbol \Phi^\text T \vec M \boldsymbol \Phi = \vec I\), where \mbox{\(\boldsymbol \Phi = [\boldsymbol \varphi_1,...,\boldsymbol \varphi_n]\)} and \(\vec I \) is the identity matrix. A reduced order model is obtained by applying the modal transformation 
\begin{equation}
 \vec q(t) = \boldsymbol \Phi_\text{p} \boldsymbol \eta_\text p(t),
\end{equation}
where $\boldsymbol \eta_\text p \in \R^{n_p}$ comprises the primary modes with \mbox{\(\boldsymbol \Phi_\text{p} = [\boldsymbol \varphi_1,...,\boldsymbol \varphi_{\text{p}}]\)} for \(n_{\text{p}} < n\). Transformation of\,\eqref{eq:system_dynamics} by left multiplying with the transposed matrix of primary eigenvectors yields the uncoupled modal equations of motion
\begin{align}
\begin{split}
	&\ddot{\boldsymbol \eta}_\text p(t) + \vec D^* \dot{\boldsymbol \eta}_\text p (t)+ \vec K^* \boldsymbol \eta_\text p(t) = \vec F_\text{a}^* \vec u(t) + \vec E_\text{d}^* \vec z(t), \hspace{1cm} t<0 \\
	&\boldsymbol \eta_\text p(0) = \boldsymbol \Phi_\text{p}^{-1} \vec q_\text{0}, \quad\dot{\boldsymbol \eta}_\text p(0) = \boldsymbol \Phi_\text{p}^{-1}\vec q_\text{1}.
\label{sec_theory_mod_red_sys}
\end{split}
\end{align}
The modal matrices are diagonal and \(\vec K^* = \diag[{\omega_1,...,\omega_p}]\), \(\vec D^* = \diag[{\zeta_1,...,\zeta_p}]\) applies, where the modal damping parameters amount to \(\zeta_i = \frac{1}{2}\Big(\frac{\alpha_0}{\omega_i} + \alpha_1\omega_i\Big)\). 

For the application of control engineering methods, the system \eqref{sec_theory_mod_red_sys} is transformed to state space by choosing the state vector as
\begin{align}
\vec x(t) = 
\begin{bmatrix}
\boldsymbol \eta_\text p(t)\\
\dot{\boldsymbol \eta}_\text p
\end{bmatrix},
\quad
\vec x(t) \in \R^{2n_\text{p}}.
\end{align}
This allows us to formulate the dynamics of the reduced order model in a state space representation
\begin{align}\label{eq:state-space}
\begin{split}
	&\dot{\vec x} = \underbrace{\begin{bmatrix} \vec 0 & \vec I \\ -\vec K^* & -\vec D^* \end{bmatrix}}_{\vec A} \vec x + \underbrace{\begin{bmatrix} \vec 0 \\ \vec F_a^* \end{bmatrix}}_{\vec B} \vec u + \underbrace{\begin{bmatrix} \vec 0 \\ \vec E_\text{d}^* \end{bmatrix}}_{\vec E} \vec z,
	\qquad t>0\\
	&\vec y = \underbrace{\begin{bmatrix} \vec C_\text t \\ \vec H_0 \end{bmatrix}}_{\vec C} \vec x, \qquad \vec x(0) = \vec x_0.
\end{split}
\end{align}
The system output $\vec y$ is comprised of the translational DOFs tracked by the camera and the forces measured by strain gauge sensors which are related to the state $\vec x$ by $\vec C_\text t$ and $\vec H_0$ respectively. Both matrices are defined in Sec.\,\ref{sec:fusion}. A total of $n_\text{p} = 10$ eigenmodes are used for state estimation in the following, which is regarded as a reasonable compromise between bandwidth, accuracy and computational load.

\section{Sensor Placement}\label{sec:sensor_placement}
With respect to more complex structures, equipping each load-bearing element with one or multiple sensors comes at a considerable cost. We are therefore interested in minimizing the number of required sensors for full-order state estimation in an optimal way. Knowledge about favorable sensor locations also helps to judge where higher redundancy, i.e. in terms of the number of sensors placed, makes sense. In\,\cite{rapp2017multimodal} and\,\cite{heidingsfeld2017gramian}, an optimality-based method for actuator and sensor placement regarding adaptive truss structures was introduced. Measures derived from the observability and controllability Gramian are employed as optimization criterion respectively. Wagner et al.\,\cite{wagner2018steady} further developed the approach for actuator placement to account for optimal steady-state disturbance compensability. As in\,\cite{rapp2017multimodal}, we consider the problem of optimal sensor placement while using the optimization method described in\,\cite{wagner2018steady}. Starting with the full sensor configuration for the system described by \eqref{eq:state-space}, we search for the subset of sensors that yields maximum observability given that one sensor is removed. Removing the strain gauge sensor with least negative impact on system observability from the full set, the procedure is repeated iteratively until the desired number of strain gauges remain. System observability is assessed by means of the trace of the observability Gramian.

\begin{figure}
 \centering
	\begin{subfigure}[c]{0.625\textwidth} 
		\centering
		\def\svgwidth{\textwidth}
		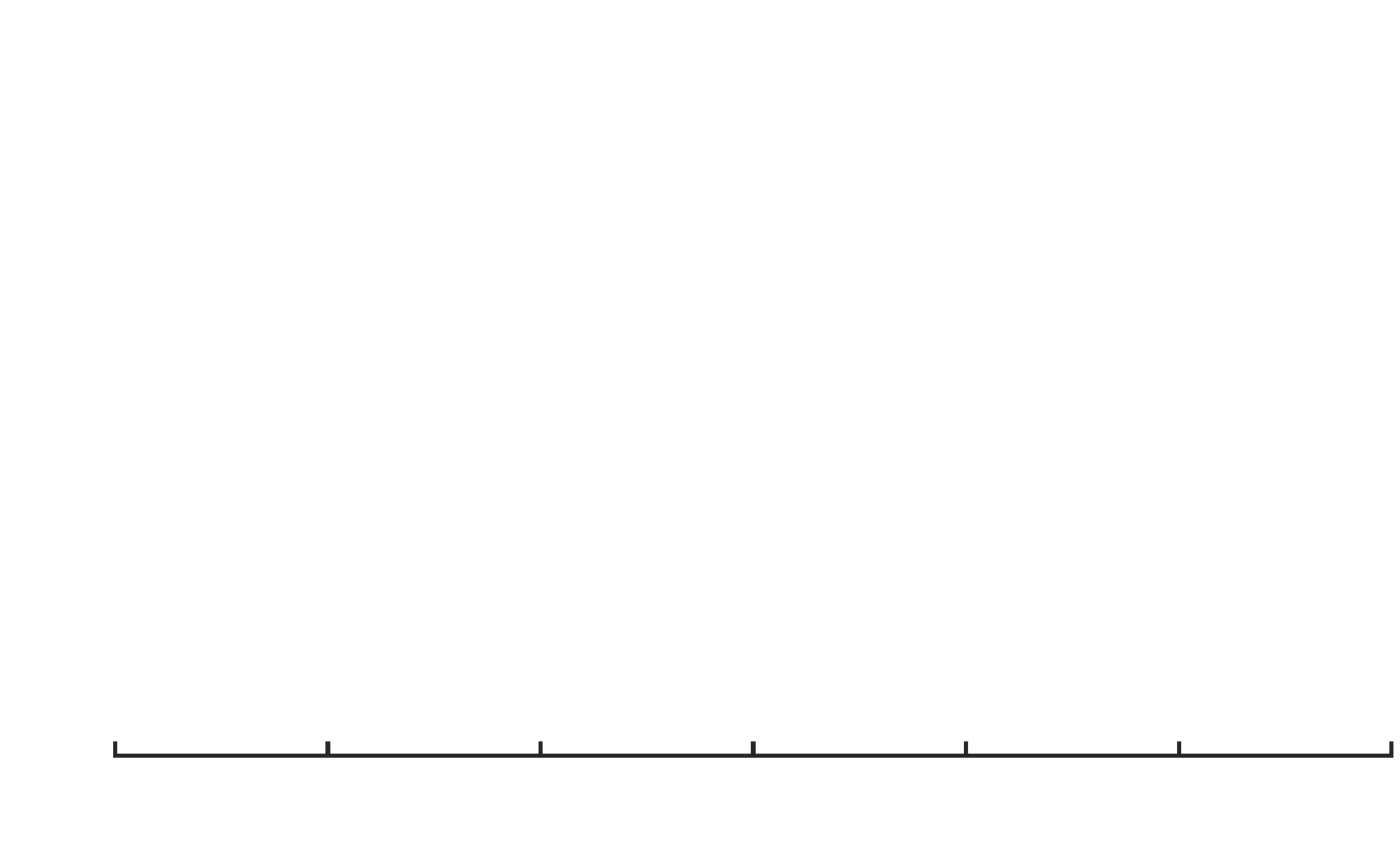 
	\end{subfigure}
	\begin{subfigure}[c]{0.3\textwidth}
		\centering
		\includegraphics{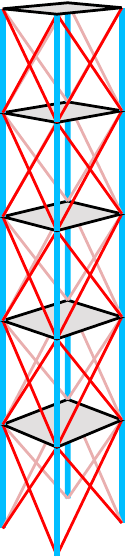}
	\end{subfigure}
 \caption[Sensor Placement Results]{Observability measure vs. the number of strain gauges used for state estimation. On the right hand side, the sensor placement for $n_\text{sg} = 40$ sensors is shown. Elements for which the axial force is measured are indicated in red. }
 \label{fig:greedy_err}
\end{figure}

Results for an optimal distribution of strain gauge sensors on the scale model are shown in Fig.\,\ref{fig:greedy_err}. The observability measure is normalized to the trace of the observability Gramian in case all sensors are used for state estimation. As expected, the observability decreases monotonically with the number of sensors $n_\text{sg}$, where the segments from $n_\text{sg} = 60$ to 40 and $n_\text{sg} = 20$ to 0 strain gauges are nearly linear. In case of the scale model, a feasible choice for the number of sensors would be at $n_\text{sg} = 40$. The negative slope approximately doubles in magnitude after this point and then continues to decrease.

To the right of the graph, the set of sensors that remains for $n_\text{sg} = 40$ is indicated in red for a schematic drawing of the structure. We observe, that no sensors remain on the columns. This is readily explained by the fact, that both active and passive columns are modeled with a lower stiffness than diagonal bracings in the example considered and are thus less sensitive to displacements. When further decreasing the number of strain gauges, sensors on the bracings of the uppermost story are completely removed for $n_\text{sg} < 28$. For $n_\text{sg} > 24$ all bracings of the lowermost story remain monitored. Thus, we assume that the lowermost bracings are more important for the observation since they also produce forces in response to displacements of higher nodes. In the same fashion, the top story bracings yield the least amount of information on the state when only bracings are considered.

In the following section, the algorithm for state estimation is described. The corresponding results are shown in Sec.\,\ref{sec:results}. Estimation accuracy with the full set of sensors is compared to the cases when only $n_\text{sg} = 40$ and 20 strain gauges are used and distributed on the structure according to the results in this section.

\section{State Estimation}\label{sec:fusion}

\begin{figure}
 \centering
 \def\svgwidth{\textwidth}
 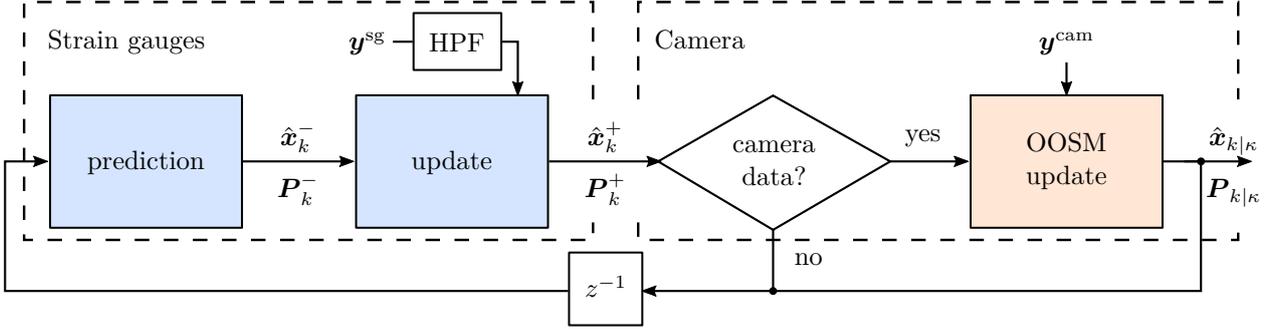 
 \caption[Filter Algorithm Block Diagram]{Sensor fusion algorithm. Strain gauge measurements are processed in each time step in the left block (blue), state estimate $\hat{\vec x}$ and estimation error covariance $\vec P$ from the previous iteration at $k-1$ are forward propagated and corrected using the strain gauge measurements $\vec y_\text{sg}$. The latter are high-pass filtered to eliminate offsets. When new camera data $\vec y_\text{cam}$ arrives, an OOSM update is performed on the current state estimate $\hat{\vec x}_k^+$ and the estimation error covariance $\vec P^+_k$.}
 \label{fig:filter_sketch}
\end{figure}

For the fusion of camera and strain gauge measurements, a discrete-time Kalman filter approach with OOSM update is used. An overview of the algorithm is shown in Fig.\,\ref{fig:filter_sketch}. Conventional prediction and update steps are performed for the strain gauge signals $\vec y^\text{sg}$ which are high-pass filtered to eliminate force offsets. The measurement update of the extended Kalman filter is used to account for discrepancies between the assumed idle state of the structure and the actual one. Image processing and the calculation of single emitter positions as described in Sec.\,\ref{sec:tracking} induce a delay on the camera measurements $\vec y^\text{cam}$. Furthermore, the camera system has a lower sampling rate than the strain gauges, for which the processing delay is considered negligible. The OOSM update accounts for both issues as it is only executed when new camera data is available and explicitly incorporates the delay. Other types of sensors sampled at different rates or even with varying delays can be readily integrated into this modular setup. In the subsequent, the processing of strain gauge data is introduced first, followed by the one-step update for the $l$-step OOSM as introduced by Bar-Shalom et al.\,\cite{bar2004one}. 

\subsection{Strain Gauge Measurements}
In each iteration of the filter algorithm, the previous state estimate $\hat{\vec x}_{k-1}^+ \in \R^{2n_\text p}$ and its associated error covariance $\vec P_{k-1}^+ \in \R^{2n_\text p\times 2n_\text p}$ are projected forward in time according to the prediction equations as stated e.\,g. in\,\cite{simon2006optimal}
\begin{align}\label{eq:predict}
	\hat{\vec x}_{k}^- &= \vec F \hat{\vec x}_{k-1}^+ \\
	\vec P_{k}^- &= \vec F \vec P_{k-1}^+ \vec F^\text T + \vec Q, \nonumber
\end{align}
where $\vec F \in \R^{2n_\text p \times 2n_\text p}$ is the discrete-time version of $\vec A$ in \eqref{eq:state-space} which is obtained according to\,\cite{simon2006optimal}. Since we deal only with the state estimation problem here, actuator inputs $\vec u$ are set to zero and the disturbances $\vec z$ are assumed to be unknown. They can be incorporated into the prediction step using the discrete time versions of $\vec B$ and $\vec E_\text d$. The matrix $\vec Q \in \R^{2n_\text p\times 2n_\text p}$ is the system noise covariance. As we assume the noise on system states to be uncorrelated, $\vec Q$ is chosen as a diagonal matrix with
\begin{equation}\label{eq:q}
	\vec Q = q \cdot \vec I
\end{equation}
The measurements $\vec y^\text{sg} \in \R^{n_sg}$ obtained from the strain gauges are related to the system state by 
\begin{equation}\label{eq:meas_sg}
	\vec y^\text{sg}_k = h_k(\vec x_k, \vec v_k) = \vec F_\text{e}(\vec x_k) + \vec v_k. 
\end{equation}
Here, $\vec v_k$ is the measurement noise and $\vec F_\text{e}$ is the corresponding vector of element forces which is calculated for each element according to
\begin{equation}\label{eq:F_e}
	F^i_\text{e}(\vec x_k) = \gamma^i \left( \left \lVert \vec d^i \vec x_k + \vec e_0^i \right \rVert_2 - L_\text{0}^{i}\right), \quad i = 1 \dots n_\text{sg}
\end{equation}
with $\gamma^i$ the spring constant, $L_0$ the element length and $\vec e_0^i$ the $i$-th element vector for $\vec x = \vec 0$. The number of strain gauges used for state estimation is given by $n_\text{sg}$. Multiplying $\vec d^i$ by $\vec x$ yields the displacement vector of the respective element. We choose not to linearize the expression \eqref{eq:meas_sg} about an equilibrium, as the initial position and deformation of the scale model varies between experiments. Also, large deformations can be achieved by exciting the XY-table, which cannot be tracked with high accuracy when using a linear Kalman filter. Given\,\eqref{eq:meas_sg} and \,\eqref{eq:F_e}, the update for force measurements is performed using the equations of the extended Kalman filter (EKF) from\,\cite{simon2006optimal}
\begin{align}
	\vec K_k &= \vec P_k^- \vec H_k^\text T\left(\vec H_k \vec P_k^-\vec H_k^\text T + \vec R_\text{sg} \right)^{-1}, \nonumber \\
	\hat{\vec x}_k^+ &= \hat{\vec x}_k^- + \vec K_k\left[\vec y_k^\text{sg} - h_k(\hat{\vec x}_k^-, \vec 0)\right], \\
	\vec P_k^+ &= \left(\vec I - \vec K_k \vec H_k \right) \vec P_k^-, \nonumber
\end{align}
with the Kalman gain $\vec K_k \in \R^{2n_\text p \times n_\text{sg}}$, the \textit{a posteriori} state estimate $\hat{\vec x}_k^+$ and error covariance $\vec P_k^+$ and 
\begin{equation} \label{eq:R_sg}
	\vec R_\text{sg} = \left(\vec b\cdot r_\text{b} + \vec c \cdot r_\text{c} \right) \cdot \vec I, \quad \vec H_k = \frac{\partial h(\vec x, \vec 0)}{\partial \vec x}\Bigg|_{\vec x = \hat{\vec x}_k^-} 
\end{equation}
where $\vec b$ is a $n_\text{sg} \times 1$ vector of ones if the corresponding elements are diagonal bracings and zeros for columns. The vector $\vec c$ is the logical inverse of $\vec b$. This way, different weights can be assigned to column and bracing forces assuming uncorrelated measurement noise. 

As the strain gauges were not calibrated for measuring absolute element forces, the predicted forces $\vec F_\text{e}$ cannot be directly related to the measured signals. To circumvent this difficulty, both the measured forces $\vec y_k^\text{sg}$ and the predicted forces are passed through a first-order Butterworth high-pass filter with a cutoff frequency of $f_\text{c} = 0.1\,\text{Hz}$. This removes the constant offset between modeled and measured element forces. Filtering the predicted measurement is not problematic in the EKF formulation since the deviations from the filtered measurements are still related to the state vector via $\vec H_k$ the same way. It comes, however, at the cost of having to rely on the model for estimating the stationary element forces. Previous calibration of each individual strain gauge sensor is recommended in order to avoid this issue.
Data from the strain gauge sensors is acquired at a sampling frequency of $f_\text{sg} = 200\,\text{Hz}$ which equals twice the maximum sampling frequency of the optical measurement system.

\subsection{Camera Measurements}
Assume a camera measurement $\vec y^\text{cam}_\kappa$ captured at time instant $t_\kappa$ with $\kappa \in [k-l,~~k-l+1)$ with $l \in \mathbb N$ is received at the current time step $k$. If the delay is lower or equivalent to one sampling time step, the method in\,\cite{bar2002update} can be used to incorporate the delayed measurement data. As the delay exceeds a single time step in our case, the one step solution for the $l$-step lag OOSM according to the algorithm $\text{B}l1$ as proposed by Bar-Shalom et al. in\,\cite{bar2004one} is employed here. The necessary steps are summarized briefly in the following. 

First of all, an equivalent measurement $Y_{k-l+1}^k = {\vec y^\text{sg}_{k-l+1}, \dots, \vec y^\text{sg}_k}$ is defined such that the sequence of measurements can be replaced by a single leap from $k-l$ to $k$. The corresponding covariance of the equivalent innovation is given by 
\begin{equation}
	\vec {S_k^*}^{-1} = \vec P_{k|k-l}^{-1} - \vec P_{k|k-l}^{-1}\left( \vec P_{k|k-l}^{-1} + {\vec P_k^+}^{-1} - \vec P_{k|k-l}^{-1} \right)^{-1} \vec P_{k|k-l}^{-1}.
\end{equation}
where $\vec P_{k|k-l}$ is computed from $\vec P_{k-l}^+$ as follows
\begin{equation}
	\vec P_{k|k-l} = \vec F_{k,k-l} \vec P_{k-l}^+ \vec F_{k,k-l}^\text T + \vec Q_{k,k-l}
\end{equation}
with the transition matrix $\vec F_{k,k-l}$ that projects the state from $k-l$ to $k$ and the associatec system noise covariance $\vec Q_{k,k-l}$. The current estimate and covariance are then retrodicted to $\kappa$ from $k$ 
\begin{align}
\begin{split}
	\hat{\vec x}_{\kappa|k} &= \vec F_{\kappa,k} \hat{\vec x}_k^+, \\
	\vec P_{\kappa|k} &= \vec F_{\kappa,k} \left( \vec P^+_k + \vec Q_{k,\kappa} - \vec P_{\vec x \vec v} - \vec P_{\vec x \vec v}^\text T \right) \vec F_{\kappa|k}^\text T, 
\end{split}
\end{align}
where $\vec F_{\kappa,k} = \vec F_{k,\kappa}^{-1}$ is the backward transition matrix and $\vec Q_{k,\kappa}$ the corresponding system noise covariance. In the above, the retrodicted noise is assumed to be zero which leads to a suboptimal, yet still accurate, solution\,\cite{bar2004one}. The retrodiction noise covariance is given by 
\begin{equation}
	\vec P_{\vec x \vec v} = \vec Q_{k,\kappa} - \vec P_{k|k-l} \vec {S_k^*}^{-1}.
\end{equation}
Given the retrodicted state and covariance, a filter gain for incorporating the OOSM is calculated as follows
\begin{equation}\label{eq:gain_oosm}
	\vec W_{k,\kappa} = \vec P_{\vec x \vec y} \left( \vec C_\text t \vec P_{\kappa|k} \vec C_\text t^\text T + \vec R_\text{cam} \right)^{-1}, 
\end{equation}
where the matrix $\vec C_\text{t}$ maps $\vec x$ to the translational DOFs tracked by the camera, $\vec R_\text{cam}$ is the camera measurement noise covariance and the covariance between the state at $k$ and the OOSM is obtained as
\begin{equation}\label{eq:p_xy}
	\vec P_{\vec x \vec y} = \left(\vec P_k^+ - \vec P_{\vec x \vec v} \right)\vec F_{\kappa,k}^\text T \vec C_\text t^\text T.
\end{equation}
Using \eqref{eq:gain_oosm} and \eqref{eq:p_xy}, the OOSM update can be formulated as
\begin{align}
\begin{split}
	\hat{\vec x}_{k|\kappa} &= \hat{\vec x}_k^+ + \vec W_{k,\kappa}\left[ \vec p_0 - \vec y_\kappa^\text{cam} - \vec C_\text t \hat{\vec x}_{\kappa|k}\right], \\
	\vec P_{k|\kappa} &= \vec P_k^+ - \vec W_{k,\kappa} \vec P_{\vec x \vec y}^\text T.
\end{split}
\end{align}
Here, $\vec p_0$ are the node coordinates for $\vec x = 0$ from which the measured nodal positions are subtracted to obtain displacements. $\vec R_\text{cam}$ in \eqref{eq:gain_oosm} is chosen as
\[
	\vec R_\text{cam} = r_\text t \cdot \vec I,
\]
assuming the measurement noise to be uncorrelated. The rank of $\vec R$ is $2 n_\text e$, where $n_\text e$ equals the number of nodal points tracked by the camera system. For simplicity, we assume in the following, that the camera delay is constant with $\kappa = k-3$ and set $\vec Q_{k,k-l} = \vec Q_{k, \kappa} = \sqrt 3 \cdot \vec Q$, where $\vec Q$ is defined by \eqref{eq:q}.

\section{Self-Tuning Algorithm}\label{sec:parameter_tuning}
With the stiffness and damping parameters initially assumed in the modeling process, state estimation results are not satisfactory. As it is usually the case, assumed model parameters do not reproduce the behavior of the real structure exactly. Neglecting friction and additional mass due to motor controllers are possible reasons for the occurrence of such mismatches. Also, table excitation is not modeled as a disturbance. Even with optimal filter coefficients, inaccurate estimations are obtained when the parameter deviations are significantly high. In the following, we present a method for simultaneous tuning of both model parameters and Kalman filter covariances based on optimization. While the parameters are varied, estimated displacements obtained using the method described in Sec.\,\ref{sec:fusion} are compared to reference measurements aiming to minimize the estimation error. In order for the algorithm to be self-tuning, selected camera measurements serve as reference measurements and are consequently not used for state estimation. Camera measurements are comparable to LDV signals in terms of precision, as visible in Fig.\,\ref{fig:quake_1}. The delay inherent in the camera signals can be compensated for by shifting the signals forward in time. Although the method presented here cannot be used for real-time adaptation of parameters, it can be applied for periodic model updating given that a number of accurate measurements are available during excitation. Considering that structural parameters are not expected to change significantly over longer periods of time, this is a reasonable option. Online joint state-parameter estimation, on the other hand, leads to a considerable increase in complexity. 

\subsection{Optimization Method}
In the optimization process, stiffness and damping parameters are adjusted along with the Kalman filter covariances in order to minimize the estimation error. To avoid calculation of the mass and stiffness matrix according to the FE model for each parameter set, only the spring constants in Eq.\,\eqref{eq:F_e} are adapted. Although less accurate, this grey-box approach ensures physical correctness and allows finding a suitable set of parameters within reasonable time. We redefine the vector of element spring constants as follows
\begin{equation}
	\boldsymbol \kappa = \vec b \cdot k_\text{b} + \vec c_\text{a} \cdot k_\text{ca} + \vec c_\text{p} \cdot k_\text{cp},
\end{equation}
where $\vec b$ is the same vector as in Eq.\,\eqref{eq:R_sg} with $k_\text{b}$ the corresponding spring constant. The column selection vector $\vec c$ is split into active columns $\vec c_\text{a}$ and passive columns $\vec c_\text{p}$. Including the Rayleigh damping coefficients and the filter covariance parameters the tunable parameters are summarized in the vector
\[
	\vec p = \begin{bmatrix} \alpha_0 & \alpha_1 & k_\text{b} & k_\text{ca} & k_\text{cp} & q & r_\text{b} & r_\text{c} & r_\text{t} \end{bmatrix}^\text T.
\]
For self-tuning, the table is excited with a chirp signal in $x$-direction. We use a disturbance input for identification on purpose since we want want to avoid active excitation of the structure in a real scenario. Considering an actual building, recorded measurements during disturbance loads such as strong winds can be used. The tuning algorithm is formulated as an optimization problem
\begin{alignat}{5}
&& \underset{\vec p}{\text{min}}
& &&&& \quad \frac{1}{n_\text{c}\cdot N} \sum_{k = k_0}^N \left | \bar{\vec y}^\text{cam}_k - \overbar{\vec C} \hat{\vec x}_k^+ \right | \\
&& \text{subject to}
& &&&& \quad \vec p^\text{l} \leq \vec p \leq \vec p^\text{u} \nonumber
\end{alignat}
where $\vec p^l$ and $\vec p^u$ define the upper and lower bounds for the parameter values. The matrix $\overbar{\vec C}$ maps the current state estimate $\hat{\vec x}_k^+$ to the $n_\text{c}$ camera measurements used as references and $N$ is the number of processed samples. To reduce errors due to deviations of emitter locations from actual nodal positions, both $\bar{\vec y}^\text{cam}$ and the estimated displacements are high-pass-filtered to obtain relative displacements for comparison. All parameters, including bounds, are normalized to the values initially assumed. The resulting stiffness parameters are adopted in our FE model and the system matrices are recalculated.

\subsection{Results for Model and Filter Parameters}\label{sec:tuning_res}
Recorded signals for a table excitation in $x$-direction with a chirp signal ranging from $0.1 - 6\,\text{Hz}$ over a period of $10\,\text{s}$ with an amplitude of $2\,\text{mm}$ are used for parameter identification. The camera measurements for the $x$-displacements of nodes 6, 13 and 22 and the $z$-displacements of nodes 5, 14 and 21 are used as references. As depicted in Fig.\,\ref{fig:msm_model}\,a), node numbers are assigned counter-clockwise and from bottom to top. Fig.\,\ref{fig:msm_model}\,b) shows which nodes are tracked by either camera, LDVs or both measurement systems. When applying the method described above, we obtain the results shown in Tab.\,\ref{tab:tuning_results}. Please note that $p^\text l$, $p^\text u$ and $p$ are normalized to the respective initial guess. Values from the spring manufacturer's data sheets were used as initial guesses for the spring constants $k_\text b$, $k_\text{ca}$ and $k_\text{cp}$. They are identical to the values used for assembling the dynamic system model before tuning. The initial guesses for other parameters reflect our beliefs about system and measurement uncertainty. More weight is put on the diagonal element force measurements as opposed to column signals because the model is less accurate for diagonal bracings. 

\begin{table}
	\parbox[t]{0.5\linewidth}{
		\caption{Parameter tuning results.}
		\label{tab:tuning_results}
		\centering
		\begin{tabular}{cc|ccc}
			Name 	& Initial guess & $p^\text l $& $p^\text u$ & $p$ \\ \hline 
			$\alpha_1$	& $0.5\times 10^{-1}$ & 0.01 & 100 & 3.43 \\ 
			$\alpha_2$	& $0.5\times 10^{-2}$ & 0.01 & 100 & 2.99 \\ 
			$k_\text{b}$	& $18200\,\text{N/m}$ & 0.5 & 2 & 1.79 \\ 
			$k_\text{ca}$	& $19500\,\text{N/m}$ & 0.5 & 2 & 0.70 \\ 
			$k_\text{cp}$	& $22100\,\text{N/m}$ & 0.5 & 2 & 1.29 \\
			$q$					& $1.0\times 10^{-7}$ & 0.01 & 100 & 3.00 \\
			$r_\text{b}$	& $1.0\times 10^{-3}$ & 0.01 & 100 & 3.14 \\ 
			$r_\text{c}$	& $1.0\times 10^{-3}$ & 0.01 & 100 & 0.26 \\ 
			$r_\text{t}$	& $1.0\times 10^{-2}$ & 0.01 & 100 & 2.46 
		\end{tabular}
	}
	\hfill
	\parbox[t]{0.45\linewidth}{
	\caption{Eigenfrequencies full-scale vs. scaled model.}
	\label{tab:eigenfrequencies}
	\centering
	\begin{tabular}{cccl}
		No. 	& Full-scale & Scaled & Mode shape \\ \hline 
		1	& $1.08$\,Hz & $0.89$\,Hz &	1st bending mode \\
		2	& $1.08$\,Hz & $0.90$\,Hz &	2nd bending mode \\
		3	& $1.84$\,Hz & $3.46$\,Hz & 1st torsion mode \\
		4	& $3.87$\,Hz & $4.86$\,Hz & 3rd bending mode \\
		5 & $3.87$\,Hz & $4.98$\,Hz & 4th bending mode \\
	\end{tabular}
	}
\end{table}

After optimization, damping coefficients and filter parameters are within reasonable ranges from their respective initial guesses. For the spring constant of both active columns $k_\text{ca}$ and passive columns $k_\text{cp}$, about 30\,\% deviation from the datasheet value is observed. Motor response to zero input $\vec u = 0$ is a possible explanation for the lower stiffness value in case of active columns. Regarding the passive columns, the increase in stiffness can be explained by friction effects that were neglected in the modeling process. There is no straightforward explanation for the $79\,\%$ increase in diagonal bracing stiffness. Slackening of diagonal cables represents a strong nonlinearity that is entirely neglected. Pretension and the associated geometric stiffness, which is different in magnitude for each individual diagonal bracing, also partially explains this result. Since table excitation is regarded as an unknown disturbance and thus not included in the model, the estimated parameters also reflect the disturbance. Furthermore, the initial pose or deformation of the scale model cannot be accurately determined with the available measurement equipment which also leads to considerable deviations between expected and measured displacements. Introduction of an additional camera or a stereo vision system for the $y$-direction is a potential remedy for the latter problem. A more accurate nonlinear model that includes the slackening effect of diagonal bracings is another possibility, however, it would lead to a significant increase in calculation time and a loss of real-time performance. The effect is also not expected to be dominant in case of a full-scale building. 

In Tab.\,\ref{tab:eigenfrequencies}, a comparison between the expected eigenfrequencies of the full-scale building and the ones obtained after tuning the stiffness parameters of the scale model is shown. Only the lowest five eigenfrequencies are listed as the mode shapes differ from the sixth mode onwards. For the demonstrator building, eigenfrequencies are calculated using a simplified FE model. The values in Tab.\,\ref{tab:eigenfrequencies} require validation after the construction of the building is finished. Similar frequencies are obtained for the first and second bending mode. The first torsion mode of the scale model is about 1.6\,Hz higher in frequency. Its eigenfrequencies are also about 1\,Hz higher for the third and fourth bending mode. With different stiffness values for active and passive columns, slight differences are observed in the bending mode frequencies due to asymmetry. Since the dynamical model of the scale model neglects some nonlinear effects, the structure's real eigenfrequencies most likely differ from the calculated ones. However, they are expected to be in a comparable range and therefore also similar to the demonstrator's eigenfrequencies.

\section{Estimation Results}\label{sec:results}

\begin{figure}
 \centering
 \def\svgwidth{0.6\textwidth}
 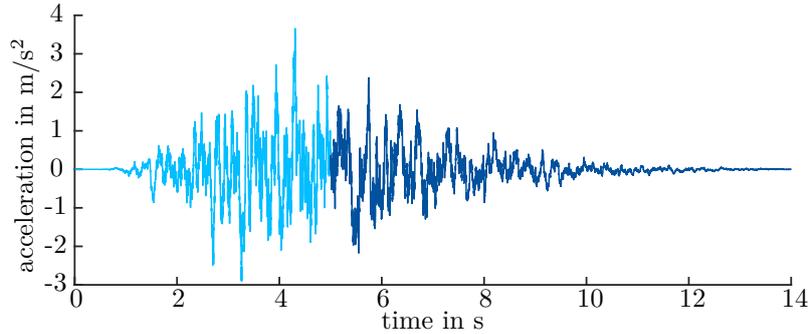 
 \caption[Earthquake excitations]{Sample generated ground acceleration record used for XY-table excitation. The displayed earthquake is of strong intensity with a peak ground acceleration of about 0.37\,g and a dominant frequency of 4\,Hz. Results in Sec.\,\ref{sec:results} are shown for the first 5\,s (highlighted).}
 \label{fig:earthquakes}
\end{figure}

Using the model and filter parameters in Tab.\,\ref{tab:tuning_results}, performance of the estimation algorithm is evaluated for earthquake-like excitations of the XY-table. For high-rise buildings, strong wind loads occur more frequently than earthquakes, but are not easily transferable to the test bench. Therefore, we choose to investigate the structure's response to earthquakes as representative dynamic loads.

The ground acceleration data of different earthquakes was generated using Matlab code from\,\cite{cheynet2019earthquake}. Based on the work of\,\cite{rofooei2001generation} and\,\cite{guo2016system}, a non-stationary Kanai-Taijimi Model is used to create artificial earthquake records. Amongst other parameters, the duration, dominant frequency as well as the standard deviation in the power spectrum can be adapted. Experiments were conducted on the test bench with various earthquakes that were downscaled in amplitude to be consistent with the scale of the model. Scaling of a ground motion record is also done in e.\,g.\,\cite{concha2016} where the earthquake is applied to a similar structure model with a height of 1.8\,m. Fig.\,\ref{fig:earthquakes} shows the ground acceleration record that is used in the following to demonstrate the performance of the state estimation method. Although several different earthquakes were applied to the structure, the conclusions that can be drawn from the results are similar, which is why only one representative ground motion is presented in detail here. The earthquake in Fig.\,\ref{fig:earthquakes} is of strong intensity with a dominant frequency of 4\,Hz and a standard deviation of 4\,Hz. The test bench table was excited with the scaled earthquake in both directions separately.

\begin{figure}
 \centering
 \def\svgwidth{0.9\textwidth}
 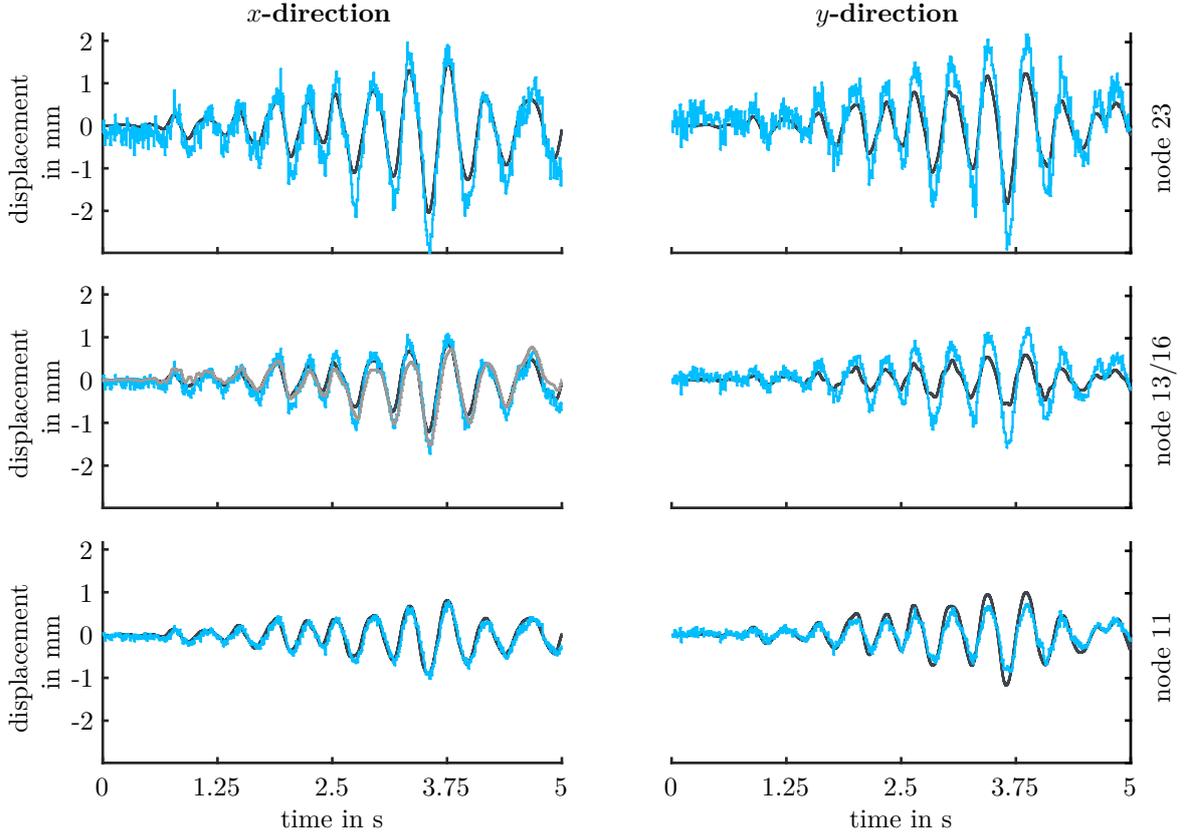 
 \caption[Estimation Results for Earthquake 1 (scaled 1:18)]{Estimated relative displacements (light blue) compared to LDV measurements (dark blue) for nodes 11, 13 $(x)$ or 16 $(y)$ and 23. The table is excited in both $x$- and $y$-direction with the earthquake (to scale) shown in Fig.\,\ref{fig:earthquakes}. The camera signal (light gray) is shown for node 13.}
 \label{fig:quake_1}
\end{figure}

Fig.\,\ref{fig:quake_1} depicts the measurement results in both $x$- and $y$-direction for a table excitation with the earthquake depicted in Fig.\,\ref{fig:earthquakes}. To enhance readability, only the onset and the first couple of seconds of the earthquake are shown. Comparative results are obtained for the remaining duration of the motion. A comparison between LDV measurements and state estimation results for the displacement of nodes 11, 13 (16 for the $y$-direction) and 23 is shown. Refer to Sec.\,\ref{sec:tuning_res} and Fig.\,\ref{fig:msm_model}\,a) for the node numbering scheme. Among the nodes tracked by LDVs, only node 13 is simultaneously tracked by the camera system, which is why the camera measurement is also visible for this node. Displacements are shown relative to the deformation before excitation and the table motion is subtracted from the LDV measurement. A maximum displacement of about 2\,mm is observed for the topmost measured node. 

Phase and amplitude of the estimated displacements in the $x$-direction are generally in good agreement with the LDV measurements, except for slight overshoots that are most pronounced for node 23 (upper left plot). When comparing the camera measurement for node 13 with the motion recorded by the LDV, sub-mm deviations as well as the processing delay are clearly visible. The latter is not always constant as it was assumed for the OOSM update in this contribution. However, the estimated displacement is mostly in phase with the reference and also tracks the amplitude more accurately. In the $y$-direction, the estimation accuracy is lower with respect to the oscillation amplitude. Overshoots are significantly higher for nodes 16 and 23 while the amplitude is underestimated for node 11. There is no visible difference in phase tracking accuracy between the $x$- and $y$-direction. When going from the lowest measured node to the highest, the signal-to-noise ratio (SNR) decreases noticeably. Taking another look at the camera measurement for node 13, we observe that its SNR is very high.

The difference between LDV and camera measurement can be attributed to two different factors. First, distortion is not completely eliminated by the projections involved in the image processing algorithm. Also, even though the LEDs were carefully positioned, their locations do not exactly match with the nodal points. Care must be taken in a real application to determine the position of LEDs as exactly as possible, as well as to ensure accurate camera calibration. With our current test bench setup, the varying camera delay cannot be determined reliably. More accurate phase tracking is expected when including this information in the OOSM update step, unleashing the full potential of the method. A straightforward explanation for the lower estimation accuracy in the $y$-direction is that the Kalman filter lacks camera measurements in this direction. Parameter identification in Sec.\,\ref{sec:parameter_tuning} was done for a motion in the $x$-direction, which is another reason for the observed differences. Supplying another camera for the $y$-direction (or using a stereo vision setup) potentially eliminates the differences. 

Since the camera sensor is observed to be very precise, we conclude that the noise on the displacement estimations results almost entirely from the strain gauge sensors which have a much lower SNR. The fact that the SNR decreases for higher stories is attributed to relative impacts of nodal displacements on each other. When a node in the first module moves, this causes relatively large displacements in the topmost story while a deformation of the top has little effect on the first floor in terms of displacements. Thus, noise of the strain gauges in the lower stories is amplified when computing the displacements of higher nodes. This has to be kept in mind when using strain gauges to estimate the building state, especially when sub-mm accuracy is to be achieved, which is desirable for a real building. Installing additional acceleration sensors can mitigate, but not fully eliminate this effect. Using more precise strain gauge sensors and amplifiers is another option. Especially when the high precision of the optical tracking system is to be fully utilized. 

\begin{figure}
 \centering
 \def\svgwidth{0.9\textwidth}
 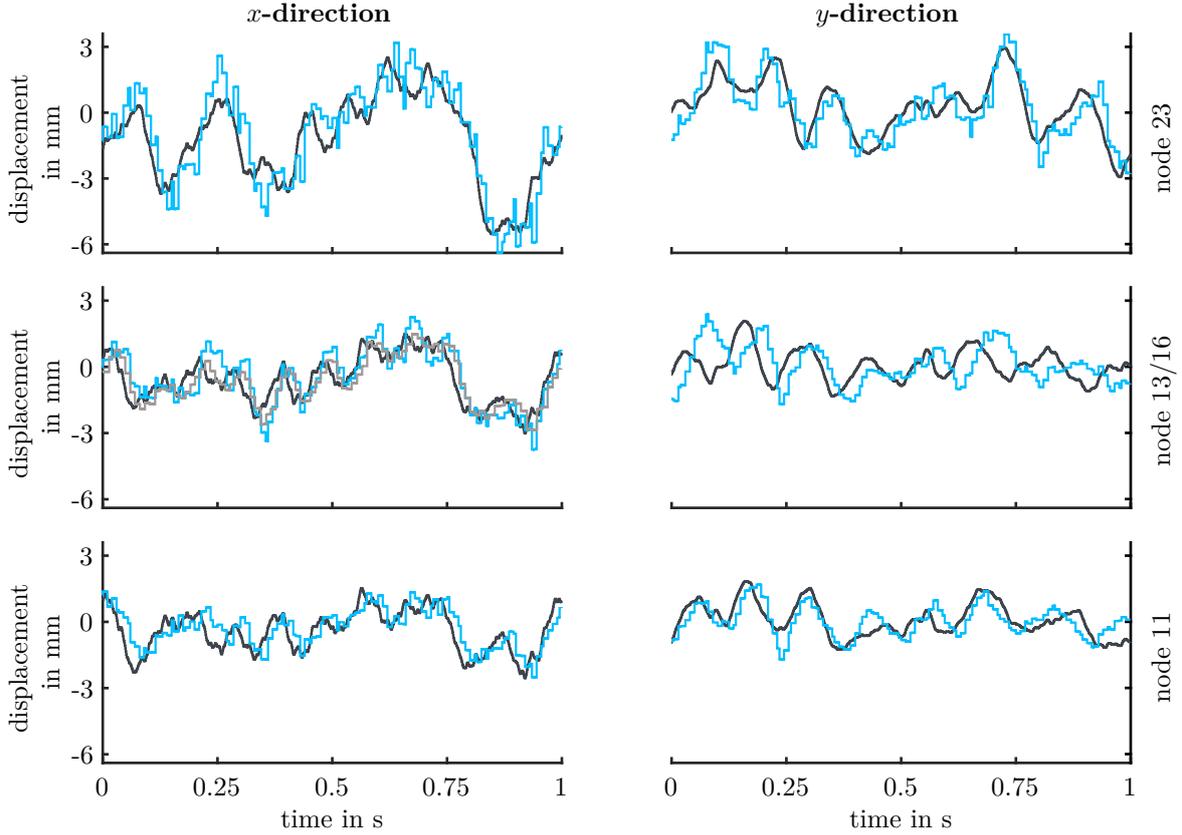 
 \caption[Estimation Results for Band-Limited White Noise]{Estimated relative displacements (light blue) compared to LDV measurements (dark blue) for nodes 11, 13 $(x)$ or 16 $(y)$ and 23. Band-limited white noise was imposed on the table's velocity controllers in both $x$- and $y$-direction separately. The camera signal (light gray) is shown for node 13.}
 \label{fig:noise}
\end{figure}

In order to explore the limits of the proposed state estimation algorithm, the velocity controllers of the test bench's XY-table were fed with band-limited white noise at a sampling frequency of 1\,kHz. Measurement results are depicted for both $x$- and $y$-direction for a representative experiment in Fig.\,\ref{fig:noise}. As opposed to Fig.\,\ref{fig:quake_1}, only a duration of one second is shown to make the comparatively high-frequency motions more discernible. 

In the $x$-direction, the estimation mostly follow the lower frequency motion recorded by the LDVs. High frequency displacements are not much discernible from noise, especially for the uppermost node. The observer does not manage to compensate the camera delay which results in a persistent phase shift. Estimation accuracy is worse in the $y$-direction, especially for node 16, where considerable phase shift is visible. 

Noise-like excitations are not to be expected in a realistic scenario. Neither the reduced order model nor the estimation algorithm are designed to operate under such conditions. It is therefore not surprising to see a drop in accuracy. In the $y$-direction, due to a lack of camera measurements, the displacement of some nodes can hardly be followed at all for high frequency motion. However, even though the excitation is exaggerated, the filter still manages to follow the low to middle frequency motion above the noise floor.

\begin{figure}
 \centering
 \def\svgwidth{0.9\textwidth}
 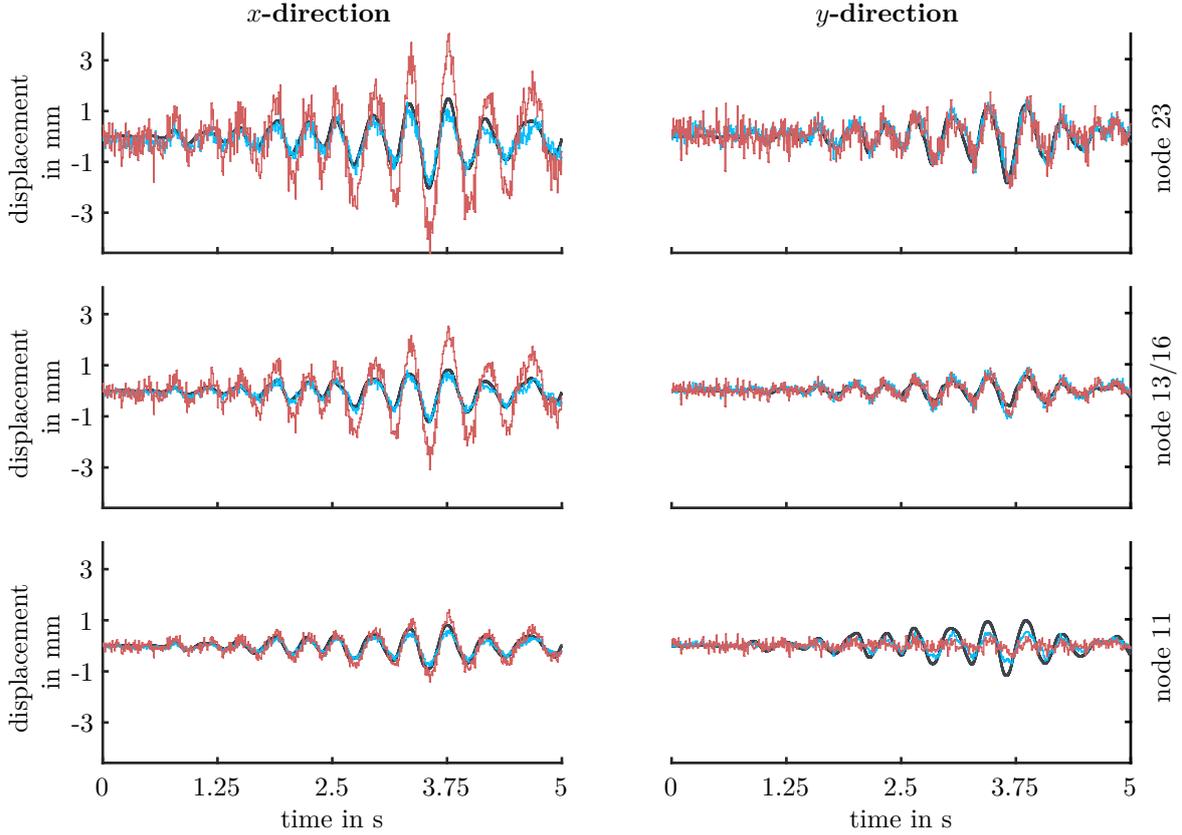 
 \caption[Estimation Results with Varying Nr. of Strain Gauges]{Estimation results for the earthquake shown in Fig.\,\ref{fig:earthquakes} when the number of strain gauges is varied. The estimated displacements for $n_\text{sg} = 40$ (light blue) and $n_\text{sg} = 20$ (red) strain gauge sensors is compared to LDV measurements (dark blue) for nodes 11, 13 $(x)$ or 16 $(y)$ and 23.}
 \label{fig:quakes_n_sg}
\end{figure}

In Sec.\,\ref{sec:sensor_placement}, optimal sensor placement on the scale model is carried out for the stiffness parameters given in Tab.\,\ref{tab:tuning_results}. In order to evaluate the effect of a reduced number of strain gauges on estimation accuracy, we conducted simulations for various values of $n_\text{sg}$. Results for $n_\text{sg} = 40$ and 20 strain gauges are shown in Fig.\,\ref{fig:quakes_n_sg}. For ease of comparison, the same excitations as in Fig.\,\ref{fig:quake_1} are used. When comparing the estimated displacements for $n_\text{sg} = 40$ to the ones visible in the previous figure, they can be seen to be almost identical. On a closer look, the amplitudes in Fig.\,\ref{fig:quakes_n_sg} are slightly lower. There is, however, no significant difference in terms of accuracy. Contrary to this observation, considerable deviations from the LDV reference in terms of both amplitude and phase are visible in case $n_\text{sg} = 20$. Especially for the excitation in $x$-direction, large estimation errors are observed. In the $y$-direction, the difference in amplitude is not so prominent, except for node 11 where the amplitude drops with the number of strain gauges decreasing. A minor increase in phase shift can also be observed. Noise on the estimated displacements increases in both directions with a reduced number of sensors.

Camera phase shift is not entirely compensated for by the OOSM update. Hence, the more strain gauge measurements are available, the better can the phase shift be corrected. Furthermore, the noise also decreases by averaging over a higher number of senors with the filter algorithm. For $n_\text{sg} = 20$ strain gauges, the observation index determined in Sec.\,\ref{sec:sensor_placement} is low when compared to $n_\text{sg} = 40$ sensors, as visible in Fig.\,\ref{fig:greedy_err}. This leads to a considerable drop in estimation quality. Readjusting the filter coefficients and stiffness parameters according to the method shown in Sec.\,\ref{sec:parameter_tuning} does not lead to significant improvements. To obtain more accurate estimation results for a low number of strain gauge sensors, a more involved system model, especially with respect to diagonal bracings, is necessary. Determination of a suitable amount of strain gauges depends on the characteristics of the system under investigation and the requirements on accuracy. Sensor numbers in the range of $90 \pm 5\,\%$ of the observability index in Fig.\,\ref{fig:greedy_err} are suggested as acceptable choices here. Applicability to other systems remains to be investigated.

\section{Conclusion and Outlook}\label{sec:conclusion}
In this contribution, we showed the application of a novel type of optical measurement system tracking emitters placed on a structure's facade for the estimation of structural dynamics. Sensor fusion with strain gauges was carried out using a Kalman filter with OOSM update for dealing with image processing delays. Since the camera measurements are very precise, they can be used for self-tuning of model and filter parameters. The proposed estimation algorithm was validated on an adaptive structures test bench with the scale 1:18 model of a lightweight adaptive high-rise. The latter was subject to scaled earthquake and random excitations for all available sensors as well as for a limited number of strain gauges selected by means of an optimal sensor placement algorithm.

Simplifications in the modeling process resulted in a discrepancy between simulated and actual structural dynamics. Although this was partially compensated for by the self-tuning algorithm, the resulting parameter mismatch was too high to be entirely justified by friction effects or the neglection of the excitation table's dynamics. Slackening of diagonal wires and the unknown initial deformation of the structure are assumed to be the main causes. In case of the actual building, both effects are less dominant. On the one hand, with the overall stiffness being significantly higher, assuming zero deformation for the initial pose is less prone to errors. On the other hand, slackening of diagonal bracings can be mostly avoided by pretensioning the elements appropriately. This shows the limits of using a scale model to study the behavior of a full-scale building. Effects that might be negligible in reality might be more dominant in a scaled version. Instead of working on a more accurate nonlinear model that is not necessary for the full-scale building, we suggest to reduce the effects by means of modifications to the test bench or the introduction of additional sensors (i.\,e. an additional camera). Care must be taken when designing a scaled version of a building in order to avoid those kind of mismatches.

Despite the aforementioned discrepancies, high estimation accuracy was achieved with the proposed sensor fusion algorithm for earthquake excitations. Camera measurements are only available in the $x$- and $z$-direction, which lead to a lower accuracy concerning $y$-displacements. Even when the test bench is excited with band-limited white noise, the motion is mostly followed, although discrepancies in phase and amplitude occurred for higher frequency displacements. When using strain gauges in a model-based estimation algorithm for tall buildings with strong physical coupling, we advice to use more accurate sensors in lower stories than in higher ones. According to the model, small deformation in the lower story trusses causes proportionally larger displacements in higher stories. Noise on strain gauges located in lower stories is thus increasingly amplified by the Kalman filter towards the top and interferes with the otherwise very precise measurements obtained from the optical system. Especially for the sensor combination considered in this contribution, this is of great importance.

In further work we aim to show control of the scale model on the test bench and to handle faults such as the occlusion of emitters. We also endeavor to decentralize the state estimation process to obtain a partially redundant system with a higher fault tolerance.

\section*{Acknowledgment}
The authors gratefully acknowledge the generous funding of this work by the German Research Foundation (DFG - Deutsche Forschungsgemeinschaft) as part of the Collaborative Research Center 1244 (SFB) "Adaptive Skins and Structures for the Built Environment of Tomorrow"/projects B02 and B04.

\bibliographystyle{elsarticle-num}

\begin{thebibliography}{10}
\expandafter\ifx\csname url\endcsname\relax
  \def\url#1{\texttt{#1}}\fi
\expandafter\ifx\csname urlprefix\endcsname\relax\def\urlprefix{URL }\fi
\expandafter\ifx\csname href\endcsname\relax
  \def\href#1#2{#2} \def\path#1{#1}\fi

\bibitem{senatore2017shape}
G.~Senatore, P.~Duffour, P.~Winslow, C.~Wise, Shape control and whole-life
  energy assessment of an ‘infinitely stiff’prototype adaptive structure,
  Smart Materials and Structures 27~(1) (2017) 015022.

\bibitem{heidingsfeld2015actuator}
M.~Heidingsfeld, E.~Arnold, C.~Tar{\'\i}n, O.~Sawodny, Actuator fault-tolerant
  control of the stuttgart smartshell, in: 2015 IEEE Conference on Control
  Applications (CCA), IEEE, 2015, pp. 996--1001.

\bibitem{aoyama2018vibration}
T.~Aoyama, L.~Li, M.~Jiang, K.~Inoue, T.~Takaki, I.~Ishii, H.~Yang, C.~Umemoto,
  H.~Matsuda, M.~Chikaraishi, et~al., Vibration sensing of a bridge model using
  a multithread active vision system, IEEE/ASME Transactions on Mechatronics
  23~(1) (2018) 179--189.

\bibitem{kang2007estimation}
L.-H. Kang, D.-K. Kim, J.-H. Han, Estimation of dynamic structural
  displacements using fiber bragg grating strain sensors, Journal of sound and
  vibration 305~(3) (2007) 534--542.

\bibitem{kim2017dynamic}
K.~Kim, H.~Sohn, Dynamic displacement estimation by fusing ldv and lidar
  measurements via smoothing based kalman filtering, Mechanical Systems and
  Signal Processing 82 (2017) 339--355.

\bibitem{park20153d}
S.~W. Park, H.~S. Park, J.~H. Kim, H.~Adeli, 3d displacement measurement model
  for health monitoring of structures using a motion capture system,
  Measurement 59 (2015) 352--362.

\bibitem{waller1990application}
H.~Waller, R.~Schmidt, The application of state observers in structural
  dynamics, Mechanical Systems and Signal Processing 4~(3) (1990) 195--213.

\bibitem{ching2006bayesian}
J.~Ching, J.~L. Beck, K.~A. Porter, Bayesian state and parameter estimation of
  uncertain dynamical systems, Probabilistic engineering mechanics 21~(1)
  (2006) 81--96.

\bibitem{smyth2007multi}
A.~Smyth, M.~Wu, Multi-rate kalman filtering for the data fusion of
  displacement and acceleration response measurements in dynamic system
  monitoring, Mechanical Systems and Signal Processing 21~(2) (2007) 706--723.

\bibitem{lourens2012joint}
E.~Lourens, C.~Papadimitriou, S.~Gillijns, E.~Reynders, G.~De~Roeck,
  G.~Lombaert, Joint input-response estimation for structural systems based on
  reduced-order models and vibration data from a limited number of sensors,
  Mechanical Systems and Signal Processing 29 (2012) 310--327.

\bibitem{azam2015dual}
S.~E. Azam, E.~Chatzi, C.~Papadimitriou, A dual kalman filter approach for
  state estimation via output-only acceleration measurements, Mechanical
  Systems and Signal Processing 60 (2015) 866--886.

\bibitem{hernandez2008state}
E.~M. Hernandez, D.~Bernal, State estimation in structural systems with model
  uncertainties, Journal of engineering mechanics 134~(3) (2008) 252--257.

\bibitem{bar2002update}
Y.~Bar-Shalom, Update with out-of-sequence measurements in tracking: exact
  solution, IEEE Transactions on aerospace and electronic systems 38~(3) (2002)
  769--777.

\bibitem{weidner2018implementation}
S.~Weidner, C.~Kelleter, P.~Sternberg, W.~Haase, F.~Geiger, T.~Burghardt,
  C.~Honold, J.~Wagner, M.~B{\"o}hm, M.~Bischoff, et~al., The implementation of
  adaptive elements into an experimental high-rise building, Steel Construction
  11~(2) (2018) 109--117.

\bibitem{opencv_library}
G.~Bradski, {The OpenCV Library}, Dr. Dobb's Journal of Software Tools, 2000.

\bibitem{ling2004element}
X.~Ling, A.~Haldar, Element level system identification with unknown input with
  rayleigh damping, Journal of Engineering Mechanics 130~(8) (2004) 877--885.

\bibitem{gawronski2004advanced}
W.~Gawronski, Advanced structural dynamics and active control of structures,
  Springer Science \& Business Media, 2004.

\bibitem{rapp2017multimodal}
P.~Rapp, M.~Heidingsfeld, M.~Böhm, O.~Sawodny, C.~Tarín, Multimodal sensor
  fusion of inertial, strain, and distance data for state estimation of
  adaptive structures using particle filtering, in: 2017 IEEE International
  Conference on Advanced Intelligent Mechatronics (AIM), 2017, pp. 921--928.

\bibitem{heidingsfeld2017gramian}
M.~Heidingsfeld, P.~Rapp, M.~B{\"o}hm, O.~Sawodny, Gramian-based actuator
  placement with spillover reduction for active damping of adaptive structures,
  in: IEEE International Conference on Advanced Intelligent Mechatronics (AIM),
  2017, IEEE, 2017, pp. 904--909.

\bibitem{wagner2018steady}
J.~L. Wagner, J.~Gade, M.~Heidingsfeld, F.~Geiger, M.~von Scheven, M.~B{\"o}hm,
  M.~Bischoff, O.~Sawodny, On steady-state disturbance compensability for
  actuator placement in adaptive structures, at-Automatisierungstechnik 66~(8)
  (2018) 591--603.

\bibitem{bar2004one}
Y.~Bar-Shalom, H.~Chen, M.~Mallick, One-step solution for the multistep
  out-of-sequence-measurement problem in tracking, IEEE Transactions on
  aerospace and electronic systems 40~(1) (2004) 27--37.

\bibitem{simon2006optimal}
D.~Simon, Optimal state estimation: Kalman, H infinity, and nonlinear
  approaches, John Wiley \& Sons, 2006.

\bibitem{cheynet2019earthquake}
E.~Cheynet,
  \href{https://www.mathworks.com/matlabcentral/fileexchange/56701-earthquake-simulation}{Earthquake
  simulation} (2019).
\newline\urlprefix\url{https://www.mathworks.com/matlabcentral/fileexchange/56701-earthquake-simulation}

\bibitem{rofooei2001generation}
F.~R. Rofooei, A.~Mobarake, G.~Ahmadi, Generation of artificial earthquake
  records with a nonstationary {K}anai--{T}ajimi model, Engineering Structures
  23~(7) (2001) 827--837.

\bibitem{guo2016system}
Y.~Guo, A.~Kareem, System identification through nonstationary data using
  time--frequency blind source separation, Journal of Sound and Vibration 371
  (2016) 110--131.

\bibitem{concha2016}
A.~Concha, L.~Alvarez-Icaza, R.~Garrido, Simultaneous parameter and state
  estimation of shear buildings, Mechanical Systems and Signal Processing 70
  (2016) 788--810.

\end{thebibliography}

\end{document}